\documentclass[aps,prl,twocolumn,showpacs]{revtex4}
\usepackage{amsfonts,amssymb,amsmath,latexsym,times} 
\usepackage[dvips]{graphics}

\begin{document}

[Phys. Rev. Lett. 102, 064101 (2009)]

\title{Universality of algebraic laws in Hamiltonian systems}

\author{Roberto Venegeroles}
\email{roberto.venegeroles@ufabc.edu.br}
\affiliation{Centro de Matem\'atica, Computa\c c\~ao e Cogni\c c\~ao, Universidade Federal do ABC, 09210-170, Santo Andr\'e, SP, Brazil}

\date{\today}

\begin{abstract}

Hamiltonian mixed systems with unbounded phase space are typically characterized by two asymptotic algebraic laws: decay of recurrence time statistics ($\gamma$) and superdiffusion ($\beta$). We conjecture the universal exponents $\gamma=\beta=3/2$ for trapping of trajectories to regular islands based on our analytical results for a wide class of area-preserving maps. For Hamiltonian mixed systems with bounded phase space the interval $3/2\leq\gamma_{b}\leq3$ was obtained, given that trapping takes place. A number of simulations and experiments by other authors give additional support to our claims.

\end{abstract}


\maketitle

Understanding the phenomenon of recurrence, introduced by Poincar\'e in the end of the 19th century \cite{Kac}, has been a matter of intense investigation over the last two decades in the theory of nonlinear dynamics. Let us consider the recurrence time probability $P(\tau)$ defined as the probability of a trajectory to return at a time $t>\tau$ to a pre-defined region. In the case of chaotic Hamiltonian systems the phenomenology of recurrence is usually divided into two groups. For {\it fully} chaotic systems its long-time decay is exponential: $P(\tau)\sim e^{-h\tau}$. This result can be understood recalling that: (i) the frequency of periodic orbits with period $\tau_{0}<\tau$ is given by the same relation, where $h$ is the Kolmogorov-Sinai entropy \cite{Sinai}, and (ii) the Bowen's theorem \cite{BM} guarantees the asymptotic equivalence of averaging over the phase space and over periodic orbits. The recurrence statistics of such systems is well understood in the literature \cite{Z}. Nonetheless, most physically realistic systems are {\it mixed}, with the phase space consisting of chaotic and regular components. For these systems sticking to regular regions takes place, and the probability $P(\tau)$ is believed to decay algebraically:
\begin{eqnarray}
P(\tau)\sim \tau^{-\gamma}.
\label{ptau}
\end{eqnarray}

The algebraic decay (\ref{ptau}) was first observed by Channon and Lebowitz \cite{CL} in the study of stochastic motion between two KAM surfaces \cite{Reichl} in the Henon quadratic map. It is known that statistics of recurrence (\ref{ptau}), clearly different from the fully chaotic one, is associated to anomalous transport \cite{ZT}, e.g., in a diffusive process where the mean square displacement increases with a power of $\tau$ not equal to $1$:
\begin{eqnarray}
\left\langle(\Delta_{\tau} x)^{2}\right\rangle\sim\tau^{\beta},
\label{mstau}
\end{eqnarray}
for an unbounded dynamical variable $x(\tau)$. This fact has motivated the use of suitable statistical theories such as continuous-time random walk (CTRW) \cite{CTRW} and fractional kinetics (FK) \cite{FK} to describe such systems. An alternative approach has been argued that the power law decay (\ref{ptau}) is due to the hierarchical structure of phase space, which can be described by Markov tree models \cite{HCM,CK}. Despite significant progress made with these theories, the understanding of algebraic decay (\ref{ptau}) from first principles, e.g., from the microscopic equations of dynamics,
is still very limited. The purpose of this Letter is to understand the microscopic mechanisms which determine the macroscopic algebraic laws (\ref{ptau}) and (\ref{mstau}) through possible connections between them.

For mixed systems the trapping of trajectories by the stable islands gives rise to power law asymptotics not only for $P(\tau)$, but also for the correlation function $C(\tau)$. Karney \cite{K} showed, using simple arguments, that these two quantities can be related through $C(\tau)\propto\sum_{t=\tau}^{\infty}P(t)$. Therefore, if $C(\tau)$ is asymptotically algebraic, then $C(\tau)\sim \tau P(\tau)$ and $P(\tau)$ satisfies Eq. (\ref{ptau}), since $\gamma>1$ guarantees the finiteness of mean recurrence times. Now, recalling the relation between the diffusion coefficient $D(\tau)$, defined by
\begin{eqnarray}
\label{diff}
D(\tau)=\frac{\left\langle (\Delta_{\tau} x)^{2}\right\rangle}{2\tau}\sim\tau^{\beta-1},
\end{eqnarray}
and the correlation function $C(\tau)$: $D(\tau)\sim\sum_{t=1}^{\tau}C(t)$ \cite{K}. If Eq. (\ref{diff}) is satisfied, then $C(\tau)$ is also asymptotically algebraic and we have the following relation
\begin{eqnarray}
\label{ab}
\gamma+\beta=3.
\end{eqnarray}
In short, by means of very simple and general arguments we show that, for sticking to regular regions in mixed systems with unbounded phase space (UPS), Eq. (\ref{ptau}) is true if, and only if, Eq. (\ref{mstau}) is also true, since $\gamma>1$ and $\beta$ satisfy the constitutive relation (\ref{ab}).

In order to find universal exponents for Hamiltonian dynamics we will consider the following class of area preserving maps
\begin{eqnarray} 
x_{t+1} = x_{t}+Kf(\theta_{t}),\qquad \theta_{t+1} = \theta_{t}+x_{t+1}\,\, \mbox{mod}\, 2\pi,
\label{map}
\end{eqnarray}
defined on the cylinder $-\pi\leq \theta\leq\pi$, $-\infty<x<\infty$, where $K$ is the stochasticity parameter. Now, let us consider general UPS Hamiltonian mixed systems. Using the relation (\ref{ab}), on basis of our analytical results for the system (\ref{map}), and from numerical simulations and experiments by other authors we will conjecture that for such systems
\begin{eqnarray} 
\gamma=\beta=3/2
\label{univt}
\end{eqnarray}
for the trapping of chaotic trajectories in the vicinity of islands of regular motion. We also confirm the Chirikov-Shepelyansky universal relation \cite{CS1}:
\begin{eqnarray}
\gamma=3,\qquad\beta=0,
\label{univkam}
\end{eqnarray}
for sticking of trajectories in the vicinity of golden invariant torii, where $\beta=0$ corresponds to the null global diffusion for $\tau\rightarrow\infty$. Eq. (\ref{univkam}) was conjectured in Ref. \cite{CS1} on basis of numerical investigations of the paradigmatic Chirikov-Taylor standard map \cite{Reichl} at the critical parameter value $K=K_{c}\approx0.97163540631$. At this value the last invariant torus is critical, e.g., susceptible to arbitrarily small perturbations. There is some controversy about the precise numerical observation of the $\gamma=3$ decay, but it is believed that this exponent should appear only for larger times \cite{MHKCS}. If we reflect on this situation through Eq. (\ref{ab}) we can understand the difficulty of achieving a conclusive numerical answer: the KAM barrier at $K=K_{c}$ is the frontier of two distinct transport regimes, namely $\beta=0$ and $\beta=1$. For the first case we have $\lim_{\tau\rightarrow\infty}D(\tau)=0$ for $K<K_{c}$, while the second exhibits normal diffusion coefficient $D\approx0.1(K-K_{c})^{3}$ for $K_{c}<K<K_{c}+0.1$ \cite{RRW}. To distinguish the two regimes on a scale of $10^{9}$ iterations for $K-K_{c}\sim10^{-11}$ as done by Weiss {\it et al}. \cite{MHKCS} it is necessary to detect $\left\langle(\Delta_{\tau} x)^{2}\right\rangle<10^{-25}$.

To verify Eq. (\ref{univt}) we will first consider the standard map, for which $f(\theta)=\sin\theta$ and several studies of its transport properties have been made \cite{Reichl}. This map exhibits, in the vicinity of $K=2l\pi$ for $l\neq0$ integer, the following power law for the diffusion coefficient:
\begin{eqnarray}
\label{dtau}
D(\tau)\sim(|l|\tau)^{1/2},
\end{eqnarray}
e.g., $\beta=3/2$, obtained analitically in a recent Letter \cite{V}. Note that this power law is valid for all $l\neq0$. We also observe that this result can be generalized for the wide class of maps (\ref{map}). If there is trapping of trajectories due to accelerator mode islands, which are created around a stable $Q$-periodic orbit $\left\{(x_{t}^{*},\theta_{t}^{*})\right\}_{1\leq t\leq Q}$, then
\begin{eqnarray}
\label{accmod}
x_{t+Q}^{*}-x_{t}^{*}=2l\pi,\qquad K\sum_{t=1}^{Q}f(\theta_{t}^{*})=2l\pi,
\end{eqnarray}
for all integer $l\neq0$. The $Q=1$ modes lead to enhanced diffusion, and the condition of stability for these is given by $|2+Kf'(\theta^{*})|<2$. Let us now assume that there are $\theta_{c}$ such that $f'(\theta_{c})=0$ and $f''(\theta_{c})\neq0$. These conditions are required to ensure the existence of a relevant asymptotic form for $\mathcal{J}_{0}(\xi)=(2\pi)^{-1}\int_{-\pi}^{\pi}d\theta\exp[i\xi f(\theta)]$:
\begin{eqnarray}
\label{Jphas}
\mathcal{J}_{0}(\xi)\sim\xi^{-1/2}\sum_{c}a_{c}\exp{\left\{i[\xi f(\theta_{c})+\pi\phi_{c}/4]\right\}},
\end{eqnarray}
where $a_{c}=(2\pi|f''(\theta_{c})|)^{-1/2}$ and $\phi_{c}=\mbox{sgn}[f''(\theta_{c})]$. Consider now $\theta^{*}=\theta_{c}+\epsilon$. We can always choose $\epsilon$ small enough such that $-4/|Kf''(\theta_{c})|<\mbox{sgn}(K)\phi_{c}\epsilon<0$, in accordance with stability condition, and $Kf(\theta_{c})\rightarrow2l\pi$ as in Eq. (\ref{accmod}) for $Q=1$, since $f(\theta_{c})\neq0$ \cite{fto}. Fortunately, by the same arguments showed in Ref. \cite{V}, we also have
\begin{eqnarray}
\label{gen}
D(\tau)\propto K\sum_{j=1}^{\tau}\mathcal{J}_{0}(jK)\sim\sum_{j=1}^{\tau}\sqrt{|l|/j}\sim(|l|\tau)^{1/2},
\end{eqnarray}
otherwise the sum in Eq. (\ref{gen}) gives terms of the type $\sum_{\eta=1}^{\tau}\exp(iq\eta)/\sqrt{\eta}$ whose sum converges by the Cauchy integral test for $\tau\rightarrow\infty$. The effective contribution of $Q\geq2$ modes for enhanced diffusion is subject to several constraints. In addition to meeting Eq. (\ref{accmod}), the stability condition of such modes follows from the trace of the monodromy matrix
\begin{eqnarray}
\label{matr}
\prod_{t=1}^{Q}\left(\begin{array}{cc}
1 & Kf'(\theta_{t}^{*})\\
1\,\,\, & 1+Kf'(\theta_{t}^{*})\end{array}\right).
\end{eqnarray}
All of these restrictions make the modes less significant as $Q$ increases. In short, the algebraic law (\ref{gen}) is closely associated with the existence of critical points $\theta_{c}$ for which the asymptotic form (\ref{Jphas}) exists. We point out a strong fact supporting the universality of $\beta=3/2$ (and thus $\gamma=3/2$): this exponent does not depend on the general form of $f(\theta)$ nor the number of critical points, just the existence of a single critical value $\theta_{c}$ is enough to justify it.

In order to reinforce the argument of universality for Eq. (\ref{univt}) we can mention a number of simulations and experiments that also point towards the universality of these exponents. Karney \cite{K} computed $P(\tau)$ for the quadratic map and Chirikov and Shepelyansky \cite{CS} did the same for the whisker map. Their numerical results agree with $\gamma=3/2$. In a recent Letter, Cristadoro and Ketzmerick \cite{CK} conjecture the universality of only Eq. (\ref{ptau}). They used a Markov tree model, originally developed by Hanson {\it et al}. \cite{HCM}, with random scaling factors for the transition probabilities. The long-time algebraic decay (\ref{ptau}) was then obtained. In order to check the universality of the Eq. (\ref{ptau}) they studied numerically an specific area preserving map of the type (\ref{map}) (for which $\theta_{c}$ exists!) for several combinations of parameters. The averaged exponent obtained was also very close to $3/2$, namely $\gamma=1.57$. There are also in the literature a number of experiments for the study of $\gamma$ and $\beta$ from the perspective of CTRW approach, for which Eq. (\ref{ab}) is also valid. Zumofen and Klafter \cite{CTRW} studied the L\'evy-walk statistics of standard map for two specific values of stochasticity parameter: $K=6.716809$ and $K=6.476939$, related to $Q=3$ and $Q=5$ accelerated modes, respectively. The exponents obtained by them are: $\gamma=1.8$ and $\beta=1.2$ for $Q=3$, and $\gamma=1.6$ and $\beta=1.4$ for $Q=5$. Note that $K=6.476939$ is closer to $K=2\pi$, for which the power law diffusion is more precise \cite{V}. This explains the best fit of the $Q=5$ mode with the universality assumption (\ref{univt}). Sanders and Larralde \cite{San} studied the ocurrence of anomalous diffusion in polygonal billiard channels. They found $\gamma=1.58$ and $\beta=1.81$ for a particular billiard model. A similar experiment was also performed by Schmiedeberg and Stark \cite{Sta} for an extended billiard model with honeycomb geometry, getting $\beta=1.72$, and by Dettmann and Cohen \cite{DC} for a wind-tree billiard model obtaining $\gamma=1.4$. We should observe that, usually, there is a coexistence between normal diffusive $\beta=1$ and superdiffusive $1<\beta<2$ processes. Thus, the algebraic laws (\ref{ptau}) and (\ref{mstau}) cannot be accounted for quantitatively by purely CTRW approach unless sufficiently large times are reached. In this sense, Gluck {\it et al}. \cite{Gluck} studied the process of chaotic diffusion for the continuous time periodic Hamiltonian $H_{\omega}(p,q,t)=p^{2}/2+\cos^{2}(q)\cos^{2}(\omega t)$, where $-\infty<q<\infty$. In order to include the normal diffusive effects they used a convolution of normal and L\'evy propagators summed over all L\'evy flight times. For $\omega=0.8$ they found $\gamma=1.55$ and $\beta=1.4$, in good agreement with Eq. (\ref{univt}). Finally, a numerical study of a stationary flow with hexagonal symmetry \cite{ZT} and two experimental studies on chaotic transport in fluid flows \cite{SWS,VAO} also agree with $\gamma\approx\beta\approx3/2$ or at least with $\beta\approx3/2$. All results described here are summarized in Table \ref{Tab1}. As we can see, the mean values obtained for $\gamma$ and $\beta$ are in agreement with the universality of Eq. (\ref{univt}).

An interesting connection between the Eq. (\ref{ab}) and the hierarchical structure of phase space can be found in a work of Zumofen {\it et al}. \cite{ZKB}. They study the motion of a particle in layered medias through its random velocity field along the $x$ direction accompanied by normal diffusional motion in the space ${\bf y}$ transverse to $x$: $x(\tau)=\int_{0}^{\tau}v[{\bf y}(t)]dt$. One of the transverse spaces studied by them is a system with hierarchically arranged connections between layers, very similar to Markov tree models used in Refs. \cite{HCM,CK}. Assuming a power law decay for the correlation of velocities $C_{v}(\tau)\sim\tau^{-\lambda}$ they conclude that $\beta=2-\min(\lambda,1)$. By Eqs. (\ref{ptau}) and $C(\tau)\sim\tau P(\tau)$ we have $\lambda=\gamma-1$, which confirms once again $\gamma+\beta=3$, since $1<\gamma,\beta<2$. They also studied another model of walks whose transverse space is one-dimensional, finding $\beta=3/2$. However, the statistical assumptions used in this model do not correspond to the dynamics of the map (\ref{map}) \cite{SL}.

\begin{table}
  \begin{center}
    \scriptsize
    \setlength{\tabcolsep}{6pt}
    \renewcommand{\arraystretch}{1.1}
    \caption{Algebraic exponents for some UPS Hamiltonian systems.}
    \vspace{2mm}
    \begin{tabular}[c]{c|c|c}
      \hline\hline
      SIMULATIONS - EXPERIMENTS & $\gamma$ & $\beta$ \\\hline
      quadratic map \cite{K,Reichl}&  $1.44$ & $-$ \\\hline
      whisker map \cite{CS,Reichl}&  $1.44$ & $-$\\\hline
      standard map for $K=6.476939$ \cite{CTRW}&  $1.6$ & $1.4$\\\hline
      parametric map model \cite{CK}&  $1.57$ & $-$\\\hline
      continuous time Hamiltonian model \cite{Gluck}&  $1.55$ & $1.4$\\\hline
      flow with hexagonal symmetry \cite{ZT}&  $-$ & $1.4$\\\hline
      two-dimensional rotating fluid flow \cite{SWS}&  $1.6$ & $1.65$\\\hline
      fluid flow with no KAM surfaces \cite{VAO}&  $-$ & $1.55$\\\hline
      zigzag billiard \cite{San}&  $1.58$ & $1.81$\\\hline
      polygonal Lorentz billiard \cite{San}&  $-$ & $1.4$\\\hline
      honeycomb billiard \cite{Sta}&  $-$ &  $1.72$\\\hline
      wind-tree billiard \cite{DC}& $-$ & $1.4$ \\\hline
      mean values & $1.54\pm0.07$ & $1.53\pm0.16$ \\\hline
    \end{tabular}\label{Tab1}
  \end{center}
\end{table}

So far, our results include only UPS Hamiltonian mixed systems, for which global diffusion takes place. Nonetheless, an additional observation on Hamiltonian mixed systems with bounded phase spaces (BPS) is in order. Suppose that for times $\tau\geq T_{0}$ the algebraic asymptotic regimes hold: $P_{u}(\tau)=C_{u}\tau^{-\gamma_{u}}$ and $P_{b}(\tau)=C_{b}\tau^{-\gamma_{b}}$, for unbounded (u) and bounded (b) systems, respectively. On the one hand, according to Poincar\'e's recurrence theorem \cite{Kac}, the return of almost all trajectories in a finite time is assured only for BPS systems. On the other hand, $\gamma_{u}=3/2>1$ guarantees the finiteness of mean recurrence times. Therefore, we also have finite return times for almost all trajectories in UPS systems. Let us focus now our attention on the trapping of trajectories by regular islands for both systems. As it is typical in such scenarios, the phase space is composed by islands of stability immersed in a sea of chaotic trajectories with no KAM surfaces. Occasionally, there may also be nonhierarchical borders between regular and chaotic regions for some types of BPS systems known to be sharply divided phase space \cite{AMK}. Now, let us construct for each arbitrary BPS system with phase space $\Omega_{b}$ its UPS periodic extension defined through translations $x(t)\rightarrow x(t)+\ell$. Here, $\ell$ is a vector of the Bravais lattice for which $\Omega_{b}$ is the first Brillouin zone and $x(t)$ becomes unrestricted. Such extension makes long recurrence times more likely. Thus, comparing the process of recurrence in both cases we must have $P_{u}(\tau)\geq P_{b}(\tau)$ which implies, through universality of Eq. (\ref{univt}), that $\gamma_{b}>\gamma_{u}=3/2$ for all $\tau\geq\max[T_{0},(C_{b}/C_{u})^{1/\gamma_{b}-3/2}]$, or $\gamma_{b}=\gamma_{u}=3/2$ if $C_{b}\leq C_{u}$. For sticking of trajectories in the vicinity of golden invariant torii a similar comparison is valid. In both cases the trajectories are equally bounded by KAM surfaces, so that we have $\gamma_{b}=\gamma_{u}=3$ by Eq. (\ref{univkam}). Thus, the maximal trapping for UPS systems should correspond to an upper limit for the maximal trapping in BPS systems. Finally, we have the inequality
\begin{eqnarray}
\label{gam2}
3/2\leq\gamma_{b}\leq3.
\end{eqnarray}
The result (\ref{gam2}) has been confirmed by numerous computer experiments \cite{AMK,Lee}. Special attention should be be payed to Hamiltonian systems with sharply divided phase space, where KAM torii coexist with ergodic components of positive measure. Altmann {\it et al}. \cite{AMK} studied numerically these systems (mushroom billiards and piecewise-linear maps) and their results suggest the universality of the exponent $\gamma_{b}=2$ in the phase space ergodic components since hierarchical borders are not present.

In conclusion, we show that the distribution of recurrence times and the power law diffusion for UPS Hamiltonian mixed systems are two macroscopic algebraic laws intimately connected through their asymptotic exponents: $\gamma+\beta=3$. This relation confirms the Chirikov and Shepelyansky conjecture of universality $\gamma=3$ for sticking of trajectories in the vicinity of golden invariant torii. We also conjecture (our main result) the universality of exponents $\gamma=\beta=3/2$ for trapping of trajectories in the vicinity of stable islands. This result was based on the analytical calculation of $\beta=3/2$ for a wide class of area preserving maps. It therefore represents a significant step towards elucidating the mechanism of the direct extraction of characteristic exponents in generic Hamiltonian systems. For trapping of trajectories to regular regions in BPS Hamiltonian mixed systems, the interval $3/2\leq\gamma_{b}\leq3$ was established, given that trapping takes place. Finally, our results agree with numerous simulations and experimental results existing in the literature. We should remark, however, that because the algebraic regime in Hamiltonian systems can take an arbitrarily long time, in general, it is not even clear for some systems whether the $\gamma=\beta=3/2$ was sufficiently reached. Another important aspect concerns the usual lack of exact power law formulas for superdiffusion coefficients: the non-acquaintance of their parametric dependence may significantly affect numerical estimates of such exponents. As previously mentioned, the $\gamma$ and $\beta$ values obtained in Ref. \cite{CTRW} for the standard map with $K=6.476939$ are closer to the foreseen values $\gamma=\beta=3/2$ than those obtained for $K=6.716809$. This occurs because the algebraic law (\ref{diff}) applies only to values of $K$ very close to $2l\pi$, since $f(\theta_{c})=1$ for such case. Now, let us return to the Gluck {\it et al} simulations \cite{Gluck}. They conclude that $\beta\approx1.4$ obtained for $\omega=0.8$ is not universal because of the value $\beta\approx1.33$ found for $\omega=0.88$. However, the Hamiltonian $H_{\omega}(p,q,t)$ studied by them can be derived from $H_{K}(p,q,t)=p^{2}/2+K\cos^{2}(q)\cos^{2}(t)$, since the following scaling transformations are made: $H\rightarrow H/\lambda$, $p\rightarrow p/\sqrt{\lambda}$, $q\rightarrow q$, $t\rightarrow\sqrt{\lambda}t$, and $K\rightarrow K$. Choosing $\lambda=K^{-1}$ we have $K=\omega^{-1/2}$ relating $H_{K}$ and $H_{\omega}$. The Hamiltonian $H_{K}$ has a similar structure to the kicked rotor Hamiltonians that lead to map (\ref{map}) \cite{nt}. Therefore, as discussed above on the standard map, it is not surprising to note some discrepancy between the two values of $\beta$ blindly fitted for the two $\omega$ values, noting that $D_{\omega}=K^{-3/2}D_{K}$ for the momentum variable $p$.

The author thanks A. Saa, W.F. Wreszinski, C.J.A. Pires, and R. Vicente for helpful discussions.


\begin{thebibliography}{99}
\bibitem{fto} If there is at lest $\theta_{c}$ such that $f(\theta_{c})=0$, Eq. (\ref{accmod}) and the stability condition are sufficient to ensure Eq. (\ref{gen}) even for $\theta^{*}\neq\theta_{c}+\epsilon$.
\bibitem{nt} The Hamiltonian $\mathcal{H}_{K}(p,q,t)=p^{2}/2-KV(q)\sum_{n}\delta(t-n)$ where $f(q)=dV(q)/dq$ leads to map (\ref{map}), where $x=p$ and $\theta=q$, for discrete time dynamics. Recalling that $\pi\delta (t-n)=1/2+\sum_{m=1}^{\infty}\cos[m(t-n)]$, a Hamiltonian very similar to $H_{K}$ can be extracted from $\mathcal{H}_{K}$ by setting $V(q)=-\frac{\pi}{2}\cos^{2}(q)$, $n=0$, and $m=2$.
\end{thebibliography}
\end{document}